\documentstyle[prl,aps,multicol,psfig]{revtex}
\begin{document}
\draft
\input epsf


\title{ Mutual Information of Sparsely Coded Associative Memory  
with Self-Control and Ternary Neurons}

\author{D. Boll\'e \thanks{Corresponding author.
 E-mail: desire.bolle@fys.kuleuven.ac.be}}
\address{Instituut voor Theoretische Fysica,
Katholieke Universiteit Leuven\\ 
Celestijnenlaan 200 D, B-3001 Leuven, Belgium }

\author{D.R.C. Dominguez }
\address{ESCET, Universidad Rey Juan Carlos,\\
C.Tulip\'an, Mostoles, 28933 Madrid, Spain, and\\
CAB (Associate NASA Astrobiology), INTA,\\
Torrejon de Ardoz, 28850 Madrid, Spain\\ }  

\author{S. Amari }
\address{RIKEN Frontier Research Program\\Wako-shi, Hirosawa 2-1\\
Saitama 351-01, Japan }

\maketitle

\begin{abstract}

\noindent
{\bf Abstract}\\

The influence of a macroscopic time-dependent threshold on the retrieval
dynamics of attractor associative memory models with ternary 
neurons $\{-1,0.+1\}$ is examined. 
If the threshold is chosen appropriately in function of the cross-talk 
noise and 
of the activity of the memorized patterns in the model, adapting itself 
in the course of the time evolution, it guarantees an autonomous 
functioning of the model.
Especially in the limit of sparse coding, it is found that this
self-control mechanism considerably improves the quality of the 
fixed-point retrieval dynamics, in particular the storage capacity, 
the basins of attraction and the information content.
The mutual information is shown to be the relevant parameter to
study the retrieval quality of such sparsely coded models.
Numerical results confirm these observations.\\

\end{abstract}

\newpage

\section*{1. Introduction}

An important property required in efficient neural network modelling is
an autonomous functioning independent from, e.g., external constraints
or control mechanisms. For fixed-point retrieval by an attractor 
associative memory model this 
requirement is mainly expressed by the robustness of its learning 
and retrieval capabilities against external noise, against
malfunctioning of some of the connections and so on. Indeed, a
model which embodies this robustness is able to perform as a
content-adressable memory having large basins of attraction for the
memorized patterns.
Intuitively one can imagine that these basins of attraction become
smaller when the storage capacity gets larger. This might occur, e.g.,
in sparsely coded models (Okada, 1996 and references cited therein).
Therefore the necessity of a control of the activity of the neurons has
been emphasized such that the latter stays the same as the activity of the
memorized patterns during the recall process.
In particular, for binary patterns with a strong bias
some external constraints were proposed on the dynamics in order to realize
this (Amit et al., 1987; Amari,1989; Buhman et al., 1989; Schwenker et
al., 1996).

An interesting question is then whether the recall process can be
optimized without imposing such external constraints, keeping the
simplicity of the (given) architecture of the network model. To this end a
self-control mechanism has been proposed in the dynamics of binary
neuron models through the
introduction of a time-dependent threshold in the transfer function. This
threshold is determined in function of both the cross-talk noise and the
activity of the memorized patterns in the network and adapts itself in
the course of the time evolution (Dominguez and Boll\'e, 1998).

The purpose of the present paper is to derive and verify this self-control
mechanism for attractor networks with multi-state neurons. There are
obvious reasons for choosing multi-state (or even analog) neurons in device
oriented applications of neural networks. To give one example, the pixels of
a colored or gray-toned pattern are represented by such neurons. In the
sequel we restrict ourselves, for convenience and without loss of
generality, to ternary neurons $\{-1,0,+1\}$.
Although the dynamics and the interesting features of the
latter have been discussed (see, e.g., Yedidia, 1989; Bouten and Engel,
1993; Boll\'e et al., 1994 and references therein) especially in the
limit of small pattern activity (Yedidia, 1989),
no activity control mechanism has been proposed in the literature until now.

The rest of this paper is organized as follows. In Section 2 we define
the attractor associative memory network model with ternary neurons and
we introduce the relevant parameters in order to discuss the quality of
the recall process. Section 3 introduces the mutual information function
for this model and discusses an explicit analytic expression for it.    
In Section 4 we study the fixed-point retrieval dynamics of the model with
complete self-control. In particular, using a probabilistic
signal-to-noise approach we obtain explicit time evolution equations
and we consider the consequences of sparse coding.
Section 5 studies the numerical solutions of this
self-controlled dynamics and compares the quality of the recall process 
in this case with the one for models without self-control. Numerical
simulations are shortly discussed. Finally, in Section 6 we present
some concluding remarks.

\section*{2. The model}

Let us consider a network with $N$ ternary neurons. At a discrete time
step $t$ the neurons $\sigma_{i,t} \in \{0,\pm 1\}, i=1,\ldots, N$ are
updated synchronously according to the rule
\begin{eqnarray}
 \sigma_{i,t+1}= F_{\theta_{t}}(h_{i,t}), \quad
 h_{i,t}= \sum_{j(\neq i)}^{N}J_{ij}\sigma_{j,t},
     \label{2.si}
\end{eqnarray}
where $h_{i,t}$ is usually called the ``local field'' (Hertz et al.,
1991) of neuron $i$ at time $t$.
In general, the transfer function $F_{\theta_{t}}$ can be a monotonic
function with  $\theta_{t}$ a time-dependent threshold parameter. Later
on it will be chosen as
\begin{equation}
     F_{\theta_t}(x) =
	\cases{
           \mbox{sign}(x)  & \mbox{if $|x|>\theta_t $}\cr
            0         &\mbox{if $|x|<\theta_t $}\, .     
	}
      \label{2.Fte}
\end{equation}
In the sequel, for theoretical simplicity in the methods used, the number of
neurons $N$ will be taken to be sufficiently large. 

The synaptic weights $J_{ij}$ are determined as a function of the memorized
patterns $\xi^{\mu}_{i}\in \{0, \pm 1\}, i=1,\ldots, N, \mu =1,\ldots, p$,
by the following learning algorithm
\begin{equation}
    J_{ij}= {C_{ij}\over Ca}NJ_{ij}^{H}, \quad
    J_{ij}^{H}= \frac{1}{N}\sum_{\mu =1}^{p} \xi^{\mu}_{i}\xi^{\mu}_{j}.
   \label{2.Ji}
\end{equation}
In this learning rule the $J_{ij}^H$ are the standard Hebb weights
(Hebb, 1949; Hertz et al., 1991) with the
ternary patterns $\xi^{\mu}_{i}$ taken to be independent identically
distributed random variables (IIDRV)
chosen according to the probability distribution
\begin{equation}
         p(\xi^{\mu}_{i})=a\delta(|\xi^{\mu}_{i}|^{2}-1)
                                        +(1-a)\delta(\xi^{\mu}_{i}).
    \label{1.px}
\end{equation}
Here $a=\langle|\xi^{\mu}_{i}|^{2}\rangle$ is the $activity$ of the
memorized patterns which is taken to be the same for all $\mu$ and
which is given by the limit $N \rightarrow \infty$ of
\begin{equation}
           a^{\mu}_{N}\equiv {1\over N}\sum_{i}|\xi_i^{\mu}|^{2}.
      \label{2.aa}
\end{equation}
The brackets $\langle \cdots \rangle$ denote the average over the memorized
patterns. The latter are unbiased and uncorrelated, i.e.,
$\langle\xi^{\mu}_{i}\rangle=0,
\langle\xi^{\mu}_{i}\xi^{\nu}_{i}\rangle=0$.
To obtain the $J_{ij}$ themselves the Hebbian weights $J_{ij}^H$
are multiplied with the $C_{ij}\in\{0,1\}$ which are chosen to be IIDRV
with probability $Pr(C_{ij}=1)=C/N, Pr(C_{ij}=C_{ji})=(C/N)^2,
C/N << 1, C>0$. This introduces the so-called extremely diluted
asymmetric architecture with $C$ measuring the average connectivity of the
network (Derrida et al., 1987). 

At this point we remark that the couplings (\ref{2.Ji}) are of
infinite range (each neuron interacts with infinitely many others) such
that our model allows a so-called mean-field theory approximation. This 
essentially means that we focus on the dynamics of a single neuron
while replacing all the other neurons by an average background local
field. In 
other words, no fluctuations of the other neurons are taken into
account, not even in response to changing the state of the chosen neuron.
In our case this approximation becomes exact because, crudely speaking, 
$h_{i,t}$ is the sum of very many terms and a central limit theorem can
be applied (Hertz et al., 1991). 

It is standard knowledge by now that synchronous mean-field theory
dynamics can be solved exactly for these diluted architectures 
(e.g., Boll\'e et al., 1994). Hence, the big advantage is that this 
will allow us to
determine the precise effects from self-control in an exact way.

In order to measure the quality of the recall process one usually
introduces the
Hamming distance between the microscopic state of the network model and
the $\mu-th$ memorized pattern, defined as
\begin{eqnarray}
  d^{\mu}_{t}\equiv
  {1\over N}\sum_{i}|\xi^{\mu}_{i}-\sigma_{i,t}|^{2}
  = a^{\mu}_N-2a^{\mu}_Nm^{\mu}_{Nt}+q_{Nt}\, .
    \label{2.Em}
\end{eqnarray}
This relation naturally leads to the definition of retrieval overlap 
between the $\mu-th$ pattern and the network state
\begin{equation}
      m^{\mu}_{N,t}\equiv {1\over Na^{\mu}_{N}}
               \sum_{i}\xi^{\mu}_{i}\sigma_{i,t},
     \label{2.Mm}
\end{equation}
and the activity of the neurons, called neural activity
\begin{equation}
           q_{N,t}\equiv {1\over N}\sum_{i}|\sigma_{i,t}|^{2}.
      \label{2.QN}
\end{equation}
The $m^{\mu}_{N,t}$ are normalized parameters within the interval
$[-1,1]$ which attain the maximal value $1$ whenever the model succeeds
in a perfect recall, i.e., $\sigma_{i,t}= \xi^{\mu}_{i}$ for all $i$.

Alternatively, the precision of the recall process can be measured by the
performance (Rieger, 1990; Shim et al., 1997)
\begin{equation}
     P^{\mu}_t \equiv \frac{1}{N}
             \sum_{i}\delta_{\xi_{i,t}^{\mu}, \sigma_{i,t}}
     \label{2.Per}
\end{equation}
which counts the relative number of correctly recalled bits. For 
subsequent manipulation, it is expedient to note that
$\delta_{\xi_{i}, \sigma_{i}}$ can be expressed as a linear combination
of terms $\xi_i^k\sigma_i^l$ with $k,l \leq 2$
\begin{equation}
          \delta_{\sigma,\xi}=
                        1-\sigma^{2}-\xi^{2}+{1\over 2}\sigma\xi+
                                      {3\over 2}\sigma^{2}\xi^{2}.
      \label{2.ds}
\end{equation}
Once the parameters (\ref{2.Mm}) and (\ref{2.QN}) are known, both these
measures for retrieval can be calculated via the dynamics (\ref{2.si}).
Here we remark that for associative memory models with neurons having
more than three states these measures for the retrieval quality can be
defined in the same way. Then, technically speaking, the performance
parameter (\ref{2.Per})-(\ref{2.ds})
will contain higher-order powers of $\sigma \xi$. 

Recently an information theoretic concept, the mutual information
(Shannon, 1948; Blahut, 1990), 
has been introduced in the study of the quality of recall of some
theoretical and practical network models  
(Dominguez and Boll\'e, 1998; Schultz and Treves, 1998; Nadal et al.,
1998 and references therein).
For sparsely coded networks in particular it turns out that this concept
is very useful 
and, in fact, to be preferred (Dominguez and Boll\'e, 1998) above the Hamming
distance.

At this point we note that it turns out to be important
to introduce the following quantity appearing in the
performance
\begin{equation}
        n_{N,t}^{\mu}\equiv
         {1\over Na^{\mu}_{N}}
	    \sum_{i}^{N}\sigma_{i,t}^{2}(\xi_{i}^{\mu})^{2}\,.
      \label{2.rN}
\end{equation}
We call this quantity the $activity$-$overlap$ since it determines
the overlap between the active neurons and the active parts of  
a memorized pattern. Although it does not play any
independent role in the time evolution of the associative memory 
model defined here it
appears explicitly in the formula for the mutual information.

\section*{3. Mutual information}

In general, in information theory the mutual information function
measures the average amount of information that can be received by
the user by observing the signal at the output of a channel (Blahut,
1990).
For the recall process of memorized patterns that we are discussing 
here, at each time step the process can be regarded as a channel with
input $\xi_i^\mu$ and output $\sigma_{i,t}$ such that this mutual
information function can be defined as (forgetting about the pattern
index $\mu$ and the time index $t$)
\begin{eqnarray}
    I(\sigma_i;\xi_i)&&=S(\sigma_i)-
           \langle S(\sigma_i|\xi_i)\rangle_{\xi_i};
             \label{3.Is} \\
    S(\sigma_i)&&\equiv -\sum_{\sigma_i}p(\sigma_i)\ln[p(\sigma_i)],
             \label{3.Ss} \\
    S(\sigma_i|\xi_i)
       &&\equiv -\sum_{\sigma_i}p(\sigma_i|\xi_i)\ln[p(\sigma_i|\xi_i)].
             \label{3.Sx}
\end{eqnarray}
Here $S(\sigma_i)$ and $S(\sigma_i|\xi_i)$ are the entropy
and the conditional entropy of the output, respectively. 
These information entropies are peculiar to the probability 
distributions of the output.
The term $\langle S(\sigma_i|\xi_i)\rangle_{\xi_i}$ is also
called the equivocation term in the recall process.
The quantity $p(\sigma_i)$ denotes the probability distribution for the
neurons at time $t$, while $p(\sigma_i|\xi_i)$ indicates the conditional
probability that the $ith$ neuron is in a state $\sigma_{i}$ at time $t$,
given that the $ith$ pixel of the memorized pattern that is being retrieved
is $\xi_{i}$.
Hereby we have assumed that the conditional probability of all the
neurons factorizes, i.e.,
 $p(\{\sigma_i\}|\{\xi_i\})=\prod_i p(\sigma_i|\xi_i)$, which is a
consequence of the mean-field theory character of our model explained in
Section~2. We remark that a similar factorization  has also been
used in Schwenker et al., 1996. 

The calculation of the different terms in the expression (\ref{3.Is})
proceeds as follows. Formally writing $\langle O \rangle
\equiv \langle O \rangle_{\sigma|\xi, \xi}=
\sum_{\xi} p(\xi) \sum_{\sigma} p(\sigma|\xi) O $ for an arbitrary
quantity $O$ the conditional probability can be obtained in a rather
straightforward way by using the complete knowledge about the system:
$\langle \xi \rangle=0, \, \langle \sigma \rangle=0, \,
\langle \sigma \xi \rangle=am, \,\langle \xi^2 \rangle=a, \,
\langle \sigma^2 \rangle=q, \, \langle \sigma^2 \xi \rangle=0, \,
\langle \sigma \xi^2 \rangle=0, \, \langle \sigma^2 \xi^2 \rangle=an,
\, \langle 1 \rangle=1$.
The result reads (we forget about the index $i$)
\begin{eqnarray}
p(\sigma|\xi)&&=
   (s_{\xi}+m\xi\sigma)\delta(\sigma^{2}-1)+
     (1-s_{\xi})\delta(\sigma),
         \nonumber\\
   s_{\xi}&&\equiv s_{0}-{q-n\over 1-a}\xi^{2},\,\,
      s_{0}\equiv {q-an\over 1-a}.
        \label{3.ps}
\end{eqnarray}
Alternatively, one can simply verify that this probability satisfies the
averages
\begin{eqnarray}
  m&&=\frac{1}{a}\langle\langle
           \sigma \xi \rangle_{\sigma|\xi}\rangle_{\xi},
      \\
  q&&=\langle\langle\sigma^{2}\rangle_{\sigma|\xi}\rangle_{\xi},
       \\
  n&&=\frac{1}{a}\langle\langle
     \sigma^{2} \xi^{2} \rangle_{\sigma|\xi}\rangle_{\xi}.
    \label{3.ms}
\end{eqnarray}
These averages are precisely equal in the limit $N \rightarrow \infty$
to the 
parameters $m$ and $q$ in (\ref{2.Mm})-(\ref{2.QN}) and to the
activity-overlap introduced in (\ref{2.rN}).
Using the probability distribution of the memorized patterns 
(\ref{1.px}), we furthermore obtain
\begin{equation}
   p(\sigma)\equiv\sum_{\xi}p(\xi)p(\sigma|\xi)=
       q\delta(\sigma^{2}-1)+(1-q)\delta(\sigma).
   \label{3.px}
\end{equation}

The expressions for the entropies defined above become
\begin{eqnarray}
S(\sigma)&=& -q \ln \frac{q}{2} - (1-q)\ln(1-q) \\
\langle S(\sigma|\xi)\rangle_{\xi}&=&
     -(1-a)[s_{0}\ln\frac{s_{0}}{2}+
           (1-s_{0})\ln(1-s_{0})] 
           \nonumber\\
 &-& a[\frac{n+m}{2} \ln\frac{n+m}{2}+ \frac{n-m}{2} \ln\frac{n-m}{2}
          \nonumber\\
       &+& (1-n)\ln(1-n)]
  \label{3.Hs}
\end{eqnarray}

These expressions are used in the next sections for discussing the
quality of the recall process of our model with self-control dynamics.

\section*{4. Self-control dynamics}

\subsection*{4.1 General equations}

It is standard knowledge (e.g., Derrida et al., 1987; Boll\'e et al., 1994)
 that the synchronous dynamics for diluted architectures 
can be solved exactly following the method based upon a signal-to-noise
analysis of the local field (\ref{2.si}) (e.g., Amari, 1977; Boll\'e et 
al., 1994; Okada, 1996 and references therein). 
Without loss of generality we focus on the recall of one pattern, say
$\mu=1$, meaning that only $m^1_{N,t}$ is macroscopic, i.e., of order $1$
and the rest of the patterns cause a cross-talk noise at each time
step of the dynamics.

Supposing that the initial state of the network model $\{\sigma_{i,0}\}$ 
is a
collection of IIDRV with mean zero and variance $q_0$ and correlated
only with memorized pattern $1$ with an overlap $m^1_0$ it is  wellknown
(see the literature cited above) that the local field
(\ref{2.si}) converges in the limit $C,N \rightarrow \infty$ to
\begin{equation}
  h_{i,0}=
      \xi^1_i m^1_0 + [\alpha q_0]^{1/2} {\cal N}(0,1)
 \label{3.hi}
\end{equation}
where the convergence is in distribution and where the quantity
${\cal N}(0,1)$ is a Gaussian random variable with mean zero and
variance unity. 
The parameters $m$ and $q$ defined in the preceding sections
have to be considered over the diluted structure and the (finite) 
loading $\alpha$ is defined by $p=\alpha C$.

This allows us to derive the first time-step in the evolution of the
network. 
For diluted architectures this first step dynamics describes the full
time evolution and we arrive at (Derrida et al., 1987; Yedidia, 1989; 
Boll\'e et al., 1993) 
\begin{eqnarray}
  m^{1}_{t+1} &=& \int^{+\infty}_{-\infty}\, Dz \frac{1}{a}
      \langle
        \xi^{1}  F_{\theta_{t}}(\xi^{1}m^{1}_{t}+
                   [\alpha q_{t}]^{1/2} \, z)\rangle
        \label{3.M1} \\
  q_{t+1} &=& \int^{+\infty}_{-\infty}\, Dz \langle
       F_{\theta_{t}}^2 (\xi^{1}m^{1}_{t}+
                  [\alpha q_{t}]^{1/2} \, z)\rangle.
      \label{3.Qt}
\end{eqnarray}
where we recall that the $\langle \cdots \rangle$ denote the average over
the distribution of the memorized patterns and $Dz=dz[\exp(-z^2/2)]/
(2\pi)^{1/2}$.

Furthermore we also have the following expression for the
activity-overlap
\begin{equation}
   n_{t+1} = \frac{1}{a}\int^{+\infty}_{-\infty}\, Dz
      \langle (\xi^1)^2
    F^2_{\theta_{t}} (\xi^{1} m^1_{t}+ [\alpha q_{t}]^{1/2} \, z)
       \rangle \, .
      \label{3.rt}
\end{equation}

For the specific transfer function defined in (\ref{2.Fte}) the
evolution equations (\ref{3.M1})-(\ref{3.Qt}) reduce to
\begin{eqnarray}
     m_{t+1} &=&
         erfc(\frac{\theta_t - m_t}{\sqrt{\alpha q_t}})
       - erfc(\frac{\theta_t + m_t}{\sqrt{\alpha q_t}}) 
        \label{G11} \\
     q_{t+1} &=& a
        [ erfc(\frac{\theta_t - m_t}{\sqrt{\alpha q_t}})
        + erfc(\frac{\theta_t + m_t}{\sqrt{\alpha q_t}}) ]
        \nonumber \\
      &+&
        2(1-a) [ erfc(\frac{\theta_t}{\sqrt{\alpha q_t}}) ] ,
    \label{G12}
\end{eqnarray}
where we have dropped the index $1$ and with the function $erfc(\cdot)$
defined as
\begin{equation}
erfc(x)\equiv
  {1\over\sqrt{2\pi}}\int_{x}^{\infty}dz \,\, e^{-z^{2}/2}\, .
        \label{4.erf}
\end{equation}
Without self-control these equations have been studied, e.g., in
Yedidia, 1989 and Boll\'e et al., 1993.

Furthermore the first term in (\ref{G12}) gives the activity-overlap.
More explicitly
\begin{equation}
   n_{t+1} =
         erfc(\frac{\theta_t - m_t}{\sqrt{\alpha q_t}})
        + erfc(\frac{\theta_t + m_t}{\sqrt{\alpha q_t}}) \, .
        \label{nfin}
\end{equation}

Of course, it is known that the quality of the recall process is
influenced by the cross-talk noise at each
time step of the dynamics. A novel idea is then to let the network
itself autonomously counter this cross-talk noise at each time step by
introducing an adaptive, hence time-dependent, threshold of the form
\begin{equation}
    \theta_{t}=c(a)\sqrt{\alpha q_{t}}\, .
     \label{2.tt}
\end{equation}
Together with Eqs.(\ref{3.M1})-(\ref{3.rt}) this relation 
describes the self-control dynamics of the network model.
For the present model with ternary neurons, this dynamical threshold is a
macroscopic parameter, thus no average must be taken over the microscopic
random variables at each time step $t$. This is different from the idea
used in some existing binary neuron models, e.g., Horn and Usher, 1989 
where a local threshold $\theta_{i,t}$ is taken
to study periodic behavior of the memorized patterns etc. Here we have 
in fact a
mapping with a threshold changing each time step, but no statistical
history intervenes in this process.

This self-control
mechanism is complete if we find a way to determine $c(a)$.
Intuitively, looking at the evolution equations 
(\ref{3.M1})-(\ref{3.Qt}) after the straightforward average over 
$\xi^1$ has been done and requiring that $m \sim 1- erfc(n_s)$ and
$q \sim a + erfc(n_s)$ with $n_s >0$ inversely proportional to the
error, 
the value of $c(a)$ should be such that mostly the argument of the 
transfer function satisfies $m-n_s\sqrt{\alpha q}\geq\theta$ and 
$n_s\sqrt{\alpha q}\leq\theta$. 
This leads to $c \sim n_s$. Here we
remark that $n_s$ itself depends on the loading $\alpha$ in the sense
that for increasing $\alpha$ it gets more difficult to have good
recall such that $n_s$ decreases. 
But it can still be chosen {\it a priori}.

\subsection*{4.2. Sparsely coded models}

In the limit of sparse coding (Willshaw et al., 1969; Palm, 1980;       
Amari, 1989; Okada, 1996 and references therein) meaning that the pattern
activity 
$a$ is very small and tends to zero for $N \rightarrow \infty$ 
it is possible to determine more precisely the factor $c(a)$
in the threshold (\ref{2.tt}). 

We start from the evolution equations (\ref{G11})-(\ref{G12})
governing the dynamics. 
To have $m\sim 1$ and $n\sim 1$  such that good
recall properties, i.e., 
$\sigma_i=\xi_i$ for most $i$ are realized, 
we want $m-\sqrt{\alpha q}\gg\theta$. 
Activity control, i.e., $q\sim a$ requires 
$\sqrt{\alpha q}\ll\theta$. 
{}From Eq.(\ref{2.tt}) we then obtain
$1\ll c(a)\ll {1\over\sqrt{\alpha a}}-1$.
Then, for $c(a)\gg 1$ the second term on the right-hand side of
Eq.(\ref{G12}) leads to
\begin{equation}
(1-a)erfc[c(a)]\to \frac{1}{c(a) \sqrt{2\pi}} \exp[{-{c(a)^{2}\over2}}]
    \label{4.pJ}
\end{equation}
This term must vanish faster than $a$ so that we obtain
$c(a)=\sqrt{-2\ln(a)}$. This turns out to be the same factor as in the
model with binary neurons (Dominguez and Boll\'e, 1998).
Very recently (Kitano and Aoyagi, 1998) such a time-dependent threshold
has also been used in binary models but for  
the recall of sparsely coded sequential patterns in the framework 
of statistical neurodynamics (Amari and Maginu, 1988; Okada, 1995)
with the assumption that the temporal correlations up to the intial time
can be neglected in the limit of an infinite number of neurons.

At this point we remark that in the limit of sparse coding,
Eqs.(\ref{G11})-(\ref{nfin}) for the overlap
and the activity-overlap become
\begin{equation}
m_{t}\sim n_{t}\sim
  1- {1\over 2} erfc({m\over\sqrt{\alpha a}}-c)]
\label{4.Mt}
\end{equation}

Using all this and technically replacing the conditions $\ll$ above by $<$,
meaning that we relax the requirement of perfect recall, we can
evaluate the critical capacity for which some small errors in the
recall process are allowed. We find
\begin{equation}
\alpha_{c}=O(|a\ln(a)|^{-1}),
    \label{4.ac}
\end{equation}
which is of the same order as the critical capacity for binary sparsely
coded network models with and without self-control (Tsodyks, 1988; Buhman
et al., 1989; Perez-Vicente, 1989; Horner, 1989; Okada, 1996; Dominguez
and Boll\'e, 1998). 

Next we turn to the quality of the recall process by the network. Because 
of the sparse coding the
Hamming distance is not a good measure since it does not distinguish
between a situation where most of the wrong neurons are inactive and a
situation where these wrong neurons are active. The errors in
recalling the active states are much more relevant since they contain
more information.
For instance, when $\sigma_{i}=0$ for all neurons the Hamming distance
$d=a$ and vanishes in the limit of sparse coding, 
while when $\sigma_{i}=1$ for all neurons it is $d=1-a$ and goes to $1$. 
Clearly, in both cases no information is transmitted. 
Furthermore, suppose that all neurons are inactive, 
i.e., $\sigma_{i}=0$, then we have that $m=0$, $q=0$ (and $n=0$), 
so the Hamming distance is $d=a$ but there is no information transmitted.
If instead of turning off the $aN$ active neurons 
(meaning that $\xi_i=\pm 1$) one would turn on $aN$ neurons among 
the inactive ones (meaning that $\xi_i= 0$) one would get 
$m=1$, $q=2a$ (and $n=1$). 
So the Hamming distance is still $d=a$, 
but now some information is transmitted. 
It is intuitively clear that the first kind of action
erases all the meaningful bits, 
while the second one does not affect essentially the code and, 
hence, leads to less important errors.

In fact, we immediately
note that for the first example $I$ is, indeed, $0$. 
In the second example we find that 
$I=-a\ln(2a)-(1-a)\ln(1-a)=S(\xi)-2a\ln(2)$ which is not much 
smaller than the entropy $S(\xi)$ of the memorized patterns.
This confirms our statement that the mutual information 
is to be preferred above the Hamming distance for discussing 
the quality of recall by sparsely coded network models.

\section*{5. Numerical Results}

Without selfcontrol the time evolution equations (\ref{G11})-(\ref{G12}) 
have been studied in Yedidia, 1989 and Boll\'e et al., 1994.
Three different types of solutions were found. 
The zero solution determined evidently by $m=0$ and $q=0$, 
a sustained activity solution defined by $m=0$ but $q \neq 0$
and solutions with both $m \neq 0$ and $q \neq 0$. 
There are both nonattracting and attracting solutions of the last type. 
It is straightforward to check that the mutual information is zero for 
both the zero solution and the sustained activity solution,
since for the dynamics considered here,
$q=n$ whenever $m=0$.
Hence we restrict ourselves to the attracting solution,
$R$ with $m > 0$.

We have solved this self-control dynamics (\ref{G11})-(\ref{G12}) for
our  model in the limit
of sparse coding and compared its recall properties with those of
non-self-controlled models.

The important features of the self-control are illustrated in Figs.~1-5.
In Fig.~1 we have plotted the information content
$I_{\alpha}\equiv pNI/\#J =\alpha I$ as a function of the threshold 
$\theta$ for $a=0.1$ and $a=0.01$ and different values of $\alpha$, 
$without$ self-control. 
This illustrates that it is rather difficult, 
especially in the limit of sparse coding, 
to choose a threshold interval such that $I_{\alpha}$ is non-zero. 
We remark that these small windows for the threshold leading to non-zero
information were used to determine what we call the optimal value 
of the threshold, $\theta_{opt}$, in Fig.~3.

In Fig.~2 we compare the time evolution of the retrieval overlap, 
$m_{t}$, starting from several initial values, $m_{0}$, 
for the  model with self-control and loading $\alpha=3$, 
an initial neural activity $q_{0}=0.01=a$ and
$\theta_{sc}=[-2(\ln a)\alpha q_t]^{1/2}$, 
with the model where the threshold is chosen by hand in an 
optimal way in the sense that we took the one giving the 
greatest information content $I_{\alpha}$ as seen in Fig.~1. 
We observe that the self-control forces more of
the overlap trajectories to go to the retrieval attractor $m=1$. 
It does improve substantially the basin of attraction. 
This is further illustrated in Fig.~3 where the basin of attraction 
for the whole retrieval phase $R$ is shown for the model with a 
$\theta_{opt}$ selected for every loading $\alpha$ and the model 
with self-control $\theta_{sc}$,
with initial value $q_{0}=0.01=a$.
We remark that even near the border of critical storage
the results are still improved. Hence the storage
capacity itself is also larger.
These results are not strongly dependent upon the initial value of 
$q_0$ as long as $q_{0}={\cal O}(a)$.

Figure 4 displays the information $I_{\alpha}$ as a function of the loading
$\alpha$ for the self-controlled model with several values of $a$.
We observe that $I_{max} \equiv I_{\alpha}(\alpha_{max})$ is reached
somewhat before the critical capacity and that it slowly increases 
with decreasing activity $a$.

This is further detailed in Fig.~5 where we have plotted $I_{max}$
and $\alpha_{max}a|\ln(a)|$
as a function of the activity on a logarithmic scale.
It shows that $I_{max}$ increases with $|\ln(a)|$ until it starts
to saturate. 
The saturation is rather slow analogously to the model with binary
neurons (Perez-Vicente 1989; Horner 1989; Dominguez and Boll\'e, 1998). 

Although we are well aware of the fact that simulations for such diluted
models are difficult because the time evolution equations have been derived
in the limit $C,N \rightarrow \infty$ with the well known
condition $C^t << N^{1/2}$ (Boll\'e et al., 1994), 
we have performed a limited number of numerical experiments. 
In this respect we note that it has recently been claimed (Arenzon and
Lemke, 1994)
in the study of binary neuron models that the analytic equation  
obtained in the extreme dilution limit also fits very well those results 
(critical storage capacity and the 
overlap as a function of the loading $\alpha$) 
obtained numerically for finite connectivity 
and under the less strong condition
$C << N$ ($C \sim 20, N\sim 16 000$)

Typical results from our simulations are shown in Fig.~6. 
There we have plotted again the information content $I_{\alpha}$ 
as a function of the loading $\alpha$
for a system with $a=0.1$, $C=100,200$ and $N=1 \times 10^6$. 
Convergence to the retrieval attractor is obtained 
after maximum ten time steps. 
We compare the analytic results with the simulations for self-control 
and without self-control with a threshold taken to be
$\theta_0=[-2(\ln a)\alpha q_0]^{1/2}$. 
Essentially, we clearly observe that self-control considerably 
improves the information content.
For fixed loading the quantitative difference
between theory and simulations is of order ${\cal O}=1/\sqrt{Ca}$. 
Clearly, further numerical work is required but this falls outside
the scope of the present study.

\section*{6. Concluding remarks}

In this paper we have introduced complete self-control in the dynamics of
associative memory networks models  with ternary neurons. 
We have studied the consequences of this self-control on the
 quality of the recall process by the network.
To this purpose we have introduced the  mutual information content for
 these models and shown that, especialy in the limit of sparse coding, 
this is a better measure for determining the quality of recall.

We find that, exactly as in the binary neuron model (with static and with
sequential patterns), 
the basins of attraction of the retrieval solutions are larger and 
the mutual information content is maximized.
We have compared the analytic results with some simulations
and we essentially confirm the improvement of the quality of recall by 
self-control.

These results strongly suggest that this idea of
self-control might be relevant for dynamical systems in general when
trying to improve the basins of attraction and  convergence times.

\section*{Acknowledgments}

We would like to thank G.~Jongen for useful discussions. 
This work has been supported by the Research Fund of the K.U.Leuven 
(grant OT/94/9).
One of us (D.B.) is indebted to the Fund for Scientific Research - Flanders
(Belgium) for financial support.

\section*{References}

\begin{description}

\item Amari S. (1977). Neural theory and association of
concept information. {\it Biological Cybernetics}, {\bf 26}, 175-185.

\item Amari S. (1989). Characteristics of sparsely encoded
associative memeory. {\it Neural Networks}, {\bf 2}, 451-457.

\item Amari S. and Maginu K. (1988). Statistical neurodynamics
of associative memory. {\it Neural Networks}, {\bf 1}, 63-73.
     
\item Amit D., Gutfreund H. and Sompolinsky H. (1987).
Information storage in neural networks with low levels of activity.
 {\it Physical Review A}, {\bf 35}, 2293- 2303.
   
\item Arenzon J.J. and Lemke N. (1994). Simulating highly
diluted neural networks. {\it  Journal of Physics A}, {\bf 27},
5161-5165.
     
\item Blahut R.E. (1990). {\it Principles and Practice of 
  Information Theory}. Reading, MA: Addison-Wesley.

\item Boll\'e D., Shim G.M., Vinck B., and Zagrebnov V.A.
(1994). Retrieval and chaos in extremely diluted Q-Ising neural
networks. {\it Journal of Statistical Physics}, {\bf 74}, 565-582.

\item Boll\'e D., Vinck B., and Zagrebnov V.A. (1993). On the
parallel dynamics of the Q-state Potts and Q-Ising neural networks.
 {\it Journal of Statistical Physics}, {\bf 70}, 1099-1119.

\item Bouten M. and Engel A. (1993). Basin of attraction in
networks of multistate neurons.  
   {\it Physical Review A}, {\bf 47}, 1397-1400.

\item Buhmann J., Divko R. and Schulten K. (1989).Associative
memory with high information content. {\it Physical Review A}, {\bf 39}, 
  2689-2692.

\item Derrida B., Gardner E., and Zippelius A. (1987). An
exactly solvable asymmetric neural network model.
    {\it Europhysics Letters}, {\bf 4}, 167-173.

\item Dominguez D.R.C. and Boll\'e D. (1998). Self-control in
sparsely coded networks.
     {\it Physical Review Letters}, {\bf 80}, 2961-2964.

\item Hebb D.O. (1949).{\it The Organization of Behavior}.
New York: John Wiley.

\item Hertz J., Krogh A. and Palmer R.G. (1991). {\it
Introduction to the Theory of Neural Computation}. Redwood City:
 Addison-Wesley  

\item Horn D. and Usher M. (1989). Neural networks with
dynamical thresholds. {\it Physical Review  A}, {\bf 40}, 1036-1044. 

\item Horner H. (1989). Neural networks with low levels of
activity: Ising vs. McCulloch-Pitts neurons. {\it  Zeitschrift f\"ur
Physik B}, {\bf 75}, 133-136.

\item Kitano K. and Aoyagi T. (1998). Retrieval dynamics of neural
networks for sparsely coded sequential patterns,
{\it Journal of Physics A}, {\bf 31}, L613-L620.

\item Nadal J-P., Brunel N. and Parga N. (1998). Nonlinear
feedforward networks with stochastic outputs: infomax implies redundancy
reduction. {\it Network: Computation in Neural Systems}, {\bf 9} 207-217.

\item Okada M. (1995). A hierarchy of macrodynamical equations
for associative memory. {\it Neural Networks}, {\bf 8}, 833-838.

\item Okada M. (1996). Notions of associative memory and sparse
coding. {\it Neural Networks}, {\bf 9}, 1429-1458 (1996).

\item Perez-Vicente C.J. (1989). Sparse coding and information
in Hebbian neural networks. {\it Europhysics Letters}, {\bf 10}, 621-625.

\item Rieger H. (1990). Storing an extensive number of
grey-toned patterns in a neural network using multi-state neurons. 
{\it Journal of Physics A},{\bf 23}, L1273-L1279.

\item Schultz S. and Treves A. (1998). Stability of the
replica-symmetric solution for the information conveyed by a neural
network. {\it Physical Review E },{\bf 57}, 3302-3310.
    
\item Shannon C.E. (1948). A mathematical theory for
communication. {\it  Bell Systems Technical Journal}, {\bf 27}, 379.

\item Shim G.M., Wong K.Y.M., and Boll\'e D. (1997). Dynamics
of temporal activity in multi-state neural networks. {\it 
    Journal of Physics A}, {\bf 30}, 2637-2652.

\item Schwenker F., Sommer F.T., and Palm G. (1996). Iterative
retrieval of sparsely coded associative memory patterns. 
    {\it Neural Networks}, {\bf 9}, 445-455.

\item Tsodyks M.V. (1988). Associative memory in asymmetric
diluted network with low level of activity.
  {\it  Europhysics Letters}, {\bf 7}, 203-208.

\item Willshaw D.J., Buneman O.P., and Longuet-Higgins H.C. (1969).
Non-holographic associative memory. {\it Nature}, {\bf 222}, 960-962.

\item Yedidia J.S. (1989). Neural networks that use three-state
neurons. {\it Journal of Physics A}, {\bf 22}, 2265-2273.

\end{description}


\newpage

\begin{figure}[t]
\begin{center}
\epsfxsize=\textwidth
\leavevmode
\epsfbox[1 1 600 800]{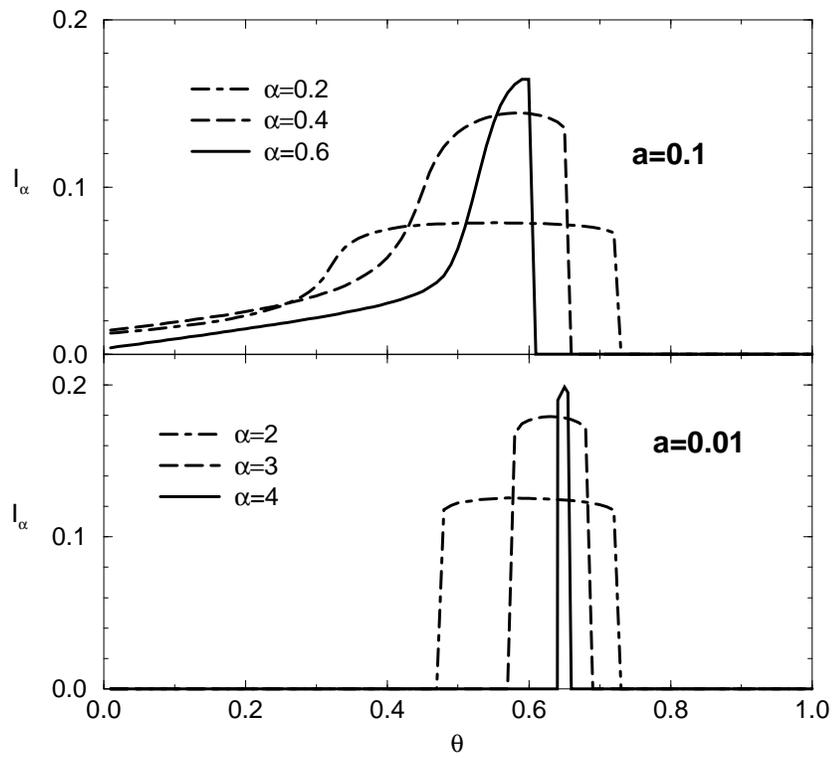}
\end{center}
\caption{ The information $I_{\alpha}$ as a function of $\theta$ without
self-control for $a=0.1$ (top) and $a=0.01$ (bottom) for several values
of $\alpha$.}
\label{i,ta}
\end{figure}

\begin{figure}[t]
\begin{center}
\epsfxsize=\textwidth
\leavevmode
\epsfbox[1 1 600 800]{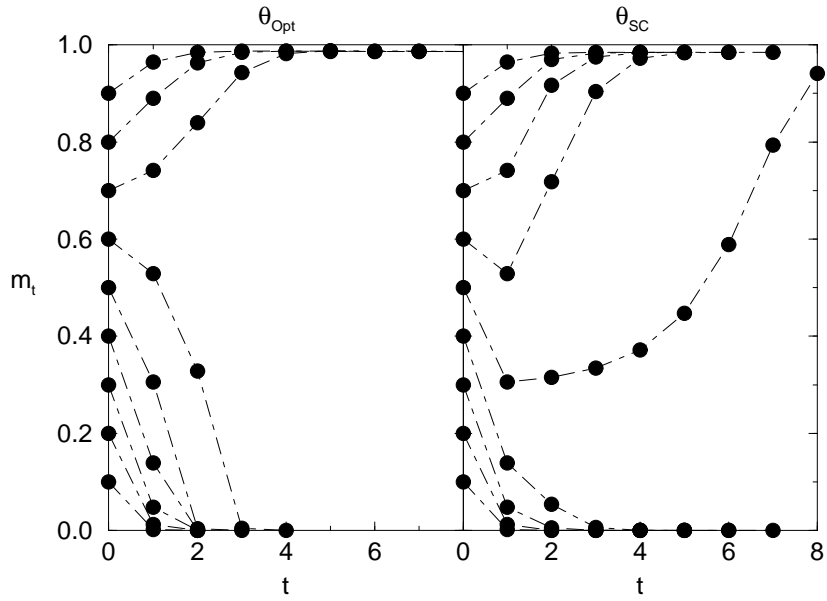}
\end{center}
\caption{ The evolution of the overlap $m_{t}$ for several initial
values $m_{0}$, with $q_{0}=0.01=a$ and $\alpha=3$ for the
model with self-control (right) and the optimal threshold model (left).
The dashed curves are a guide to the eye.}
\label{m,ta}
\end{figure}

\begin{figure}[t]
\begin{center}
\epsfxsize=\textwidth
\leavevmode
\epsfbox[1 1 600 800]{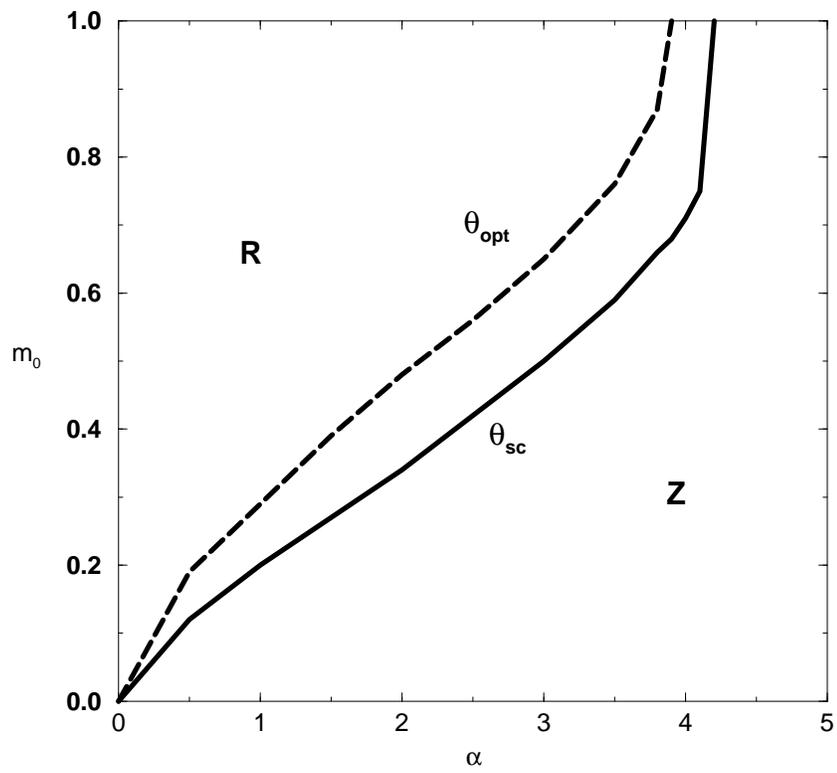}
\end{center}
\caption{ The basin of attraction as a function of $\alpha$ for
$a=0.01$ and initial $q_{0}=a$ for the self-controlled model
(full line) and the optimal threshold model (dashed line). }
\label{m0,a}
\end{figure}

\begin{figure}[t]
\begin{center}
\epsfxsize=\textwidth
\leavevmode
\epsfbox[1 1 600 800]{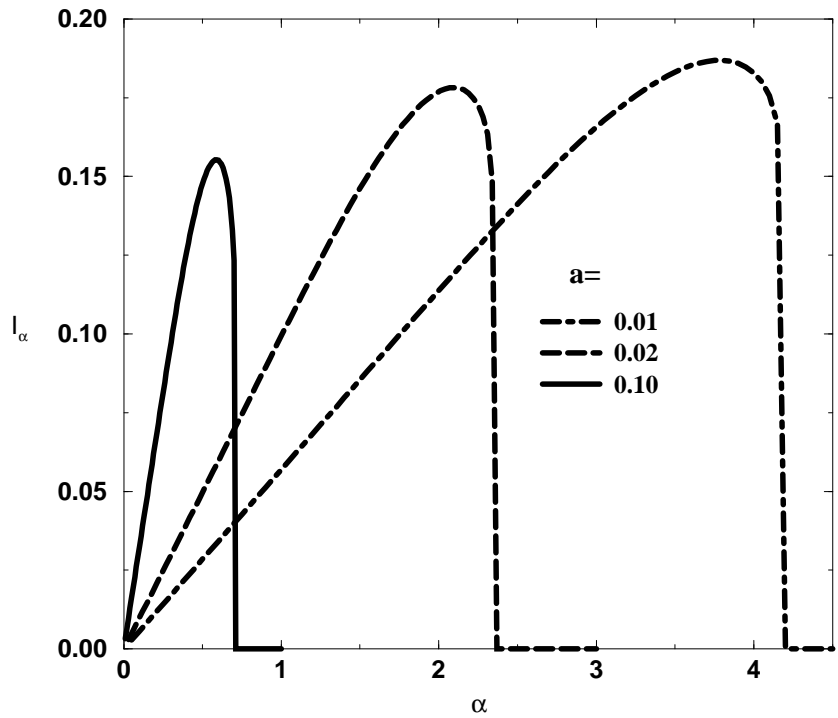}
\end{center}
\caption{ The information $I_{\alpha}$ as a function of $\alpha$ for the
self-controlled model with several values of~$a$. }
\label{i,aA}
\end{figure}

\begin{figure}[t]
\begin{center}
\epsfxsize=\textwidth
\leavevmode
\epsfbox[1 1 600 800]{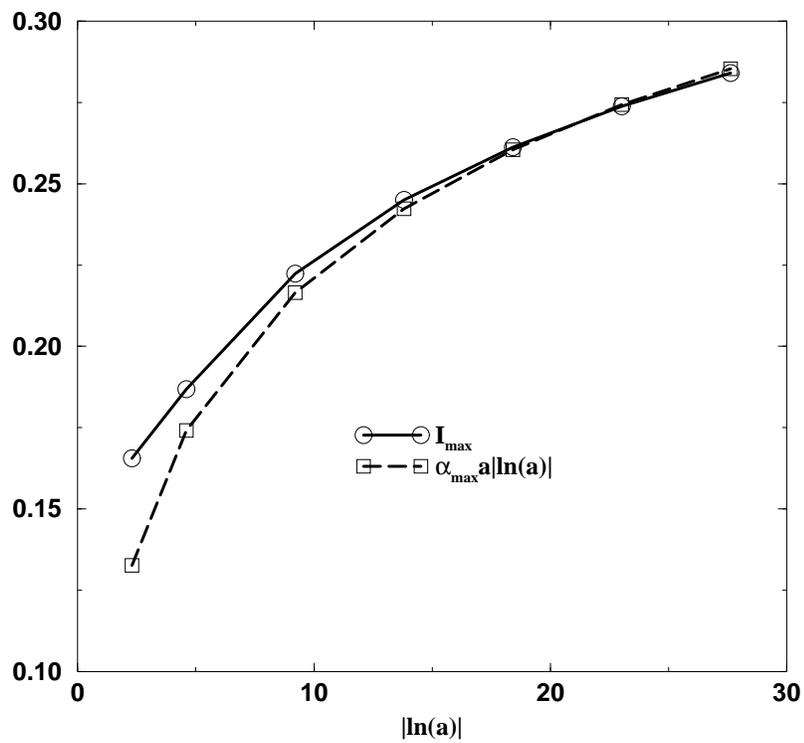}
\end{center}
\caption{ The maximal information $I_{max}$ and
$\alpha_{max} a|\ln(a)|$ as a function of $|\ln(a)|$. }
\label{ia,A}
\end{figure}

\begin{figure}[t]
\begin{center}
\epsfxsize=\textwidth
\leavevmode
\epsfbox[1 1 600 800]{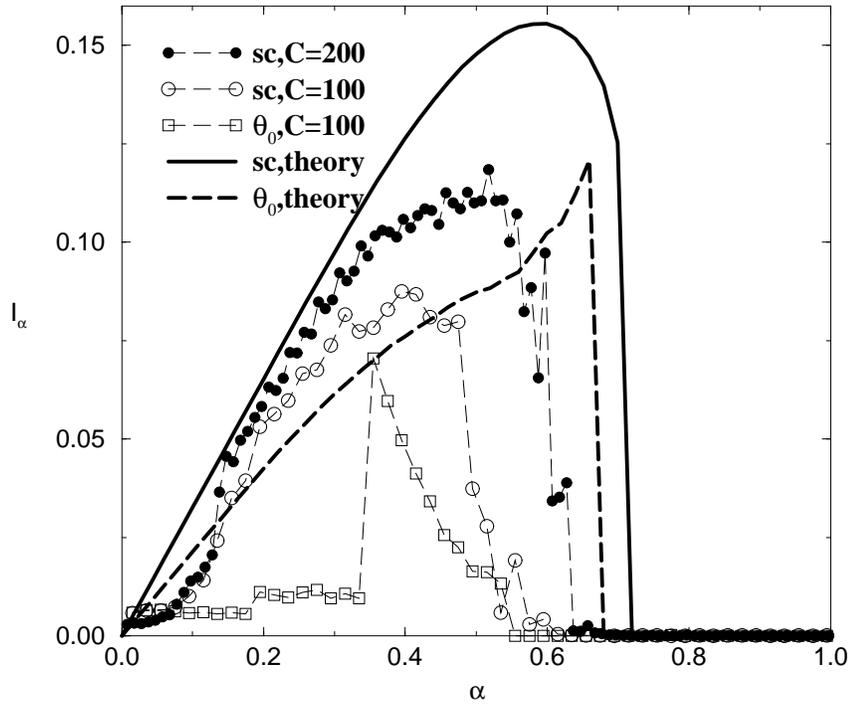}
\end{center}
\caption{ The information $I_{\alpha}$ as a function of the loading $\alpha$
for $a=0.1$ and initial value $q_0=a$ with self-control and with
fixed threshold $\theta_0$.
Analytic results are compared with simulations for $N=10^6$ and 
$C=100, 200$.}
\label{i,aS}
\end{figure}

\end{document}